\documentstyle[prl,aps,multicol,epsf]{revtex}
\begin{document}
\draft
\title{Small Angle Shubnikov-de Haas Measurements in Silicon MOSFET's: the
Effect of Strong In-Plane Magnetic Field.}
\author{Sergey~A.~Vitkalov, Hairong Zheng, K. M. Mertes and M.~P.~Sarachik}
\address{Physics Department, City College of the City University of New York,
New York, New York 10031}
\author{T.~M.~Klapwijk}
\address{Delft University of Technology, Department of Applied Physics,
2628 CJ
Delft, The Netherlands}
\date{\today}
\maketitle

\begin{abstract}
Measurements in magnetic fields applied at small angles relative to the
electron plane in silicon MOSFETs indicate a factor of two increase of
the frequency of Shubnikov-de Haas oscillations at $H>H_{sat}$.  This signals
the
onset of full spin polarization above $H_{sat}$, the parallel field
above which the resistivity
saturates to a constant value.  For $H<H_{sat}$,
the phase of the second harmonic of the oscillations relative to the first
is consistent with scattering
events that depend on the overlap instead of the sum of the
spin-up
and spin-down densities of states.

\end{abstract}


\begin{multicols}{2}

A great deal of interest has recently been focussed on the anomalous
behavior of two-dimensional (2D) systems of electrons\cite{krav,popovic}
and holes\cite{coleridge,shahar,cambridge} whose resistivities unexpectedly
decrease with decreasing temperature, behavior that is generally associated
with metals rather than insulators\cite{NAS}.
One of the most intriguing characteristics of these systems is their enormous
response to magnetic fields applied in the plane of the
electrons\cite{simonian,pudalov,dolgopolov} or holes\cite{cambridge,yoon}:
the resistivity increases sharply by more than an order of magnitude,
saturating
to a constant plateau value above a magnetic field $H_{sat}$.

In this paper we report studies of the resistivity
of the 2D electron system in silicon MOSFETs in magnetic fields
applied at small angles $\phi$ with respect to the plane.  This
allows a study of the Shubnikov-de Haas (SdH) oscillations
in perpendicular fields sufficiently small that the orbital motion has a
negligible effect on the response to the in-plane component of the
magnetic field.  At
small
tilt angles $\phi$, the SdH oscillations plotted versus filling factor
have twice the period below $H_{sat}$ compared with the period above
$H_{sat}$.  This
implies that the electron system is fully spin-polarized in high fields,
$H>H_{sat}$, where
the resistivity has reached saturation.  Detailed examination of the
oscillations in fields below $H_{sat}$ indicates unusual behavior
consistent with electron scattering that depends on the
product rather
than the sum of the spin-up and spin-down densities of states.

Two silicon MOSFETs with mobilities $\mu \approx 20,000\;$V/cm$^2$ s
at $T=4.2$ K were used in these studies.  Contact resistances were
minimized by
using samples with a split-gate geometry, which permit high
densities in the
vicinity of the contacts while allowing independent control of the density of
the 2D system under investigation.  Standard $AC$ four-probe techniques were
used at $3$ Hz to measure the resistance in the linear regime using 
currents typically below $5$ nA.  
Data were taken on samples mounted on a rotating platform
in a $^3$He Oxford Heliox system at
temperatures down to $0.235$ K in magnetic fields up to $12$ T.

\vbox{
\vspace{0.2in}
\hbox{
\hspace{-0.2in} 
\epsfxsize 3.3in \epsfbox{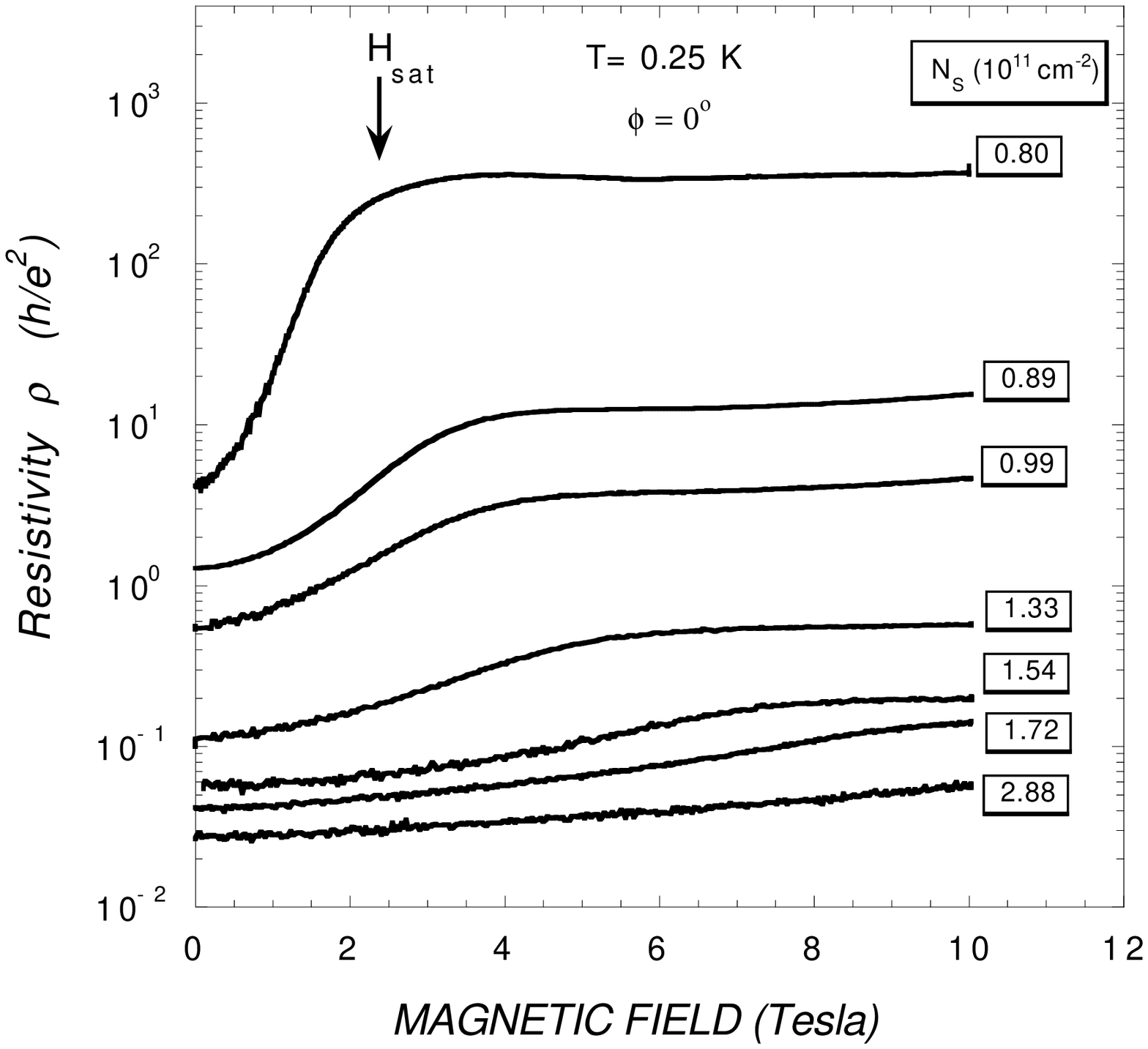} 
}
}
\refstepcounter{figure}
\parbox[b]{3.3in}{\baselineskip=12pt \egtrm FIG.~\thefigure.
Resistivity of the 2D electron system in silicon MOSFETs
versus in-plane magnetic field for different
densities, as labeled.  Data
are shown at $T=245$ mK.  The arrow indicates $H_{sat}$ for
electron density $n_s=0.80 \times 10^{11}$ cm$^{-2}$.
\vspace{0.10in}
}
\label{1}

Measurements were first taken with the plane of the sample oriented parallel
to the magnetic field\cite{alignment}.
The resistance, $R_{xx}$, is shown in Fig. 1 as a function of field
for different fixed gate voltages spanning densities between $n_s = 0.8
\times 10^{11}$ cm$^{-2}$ and $n_s = 2.88 \times 10^{12}$  cm$^{-2}$ (the
zero field critical density for the metal-insulator transition is
$n_c \approx 0.84 \times 10^{11}$ cm$^{-2}$).    Consistent with earlier
findings
\cite{simonian,pudalov,yoon,magneto,okamoto}, the in-plane
($\phi=0$) magnetoresistance rises dramatically with increasing field and
saturates above a density-dependent field $H_{sat}(n_s)$\cite{defineH}.

The sample was then rotated to make a small angle $\phi$  with respect
to the field, so that the in-plane
component was almost equal to the total field $H_{\Vert} \approx H$, while
the projection in the perpendicular direction, $H_{\perp} \approx \phi H$,
remained relatively small even in high fields.  $R_{xx}$ and $R_{xy}$ were
measured simultaneously as a function of magnetic field for fixed angle
$\phi$,
temperature T, and density $n_s$.

\vbox{
\vspace{0.2in}
\hbox{
\hspace{-0.3in} 
\epsfxsize 3.3in \epsfbox{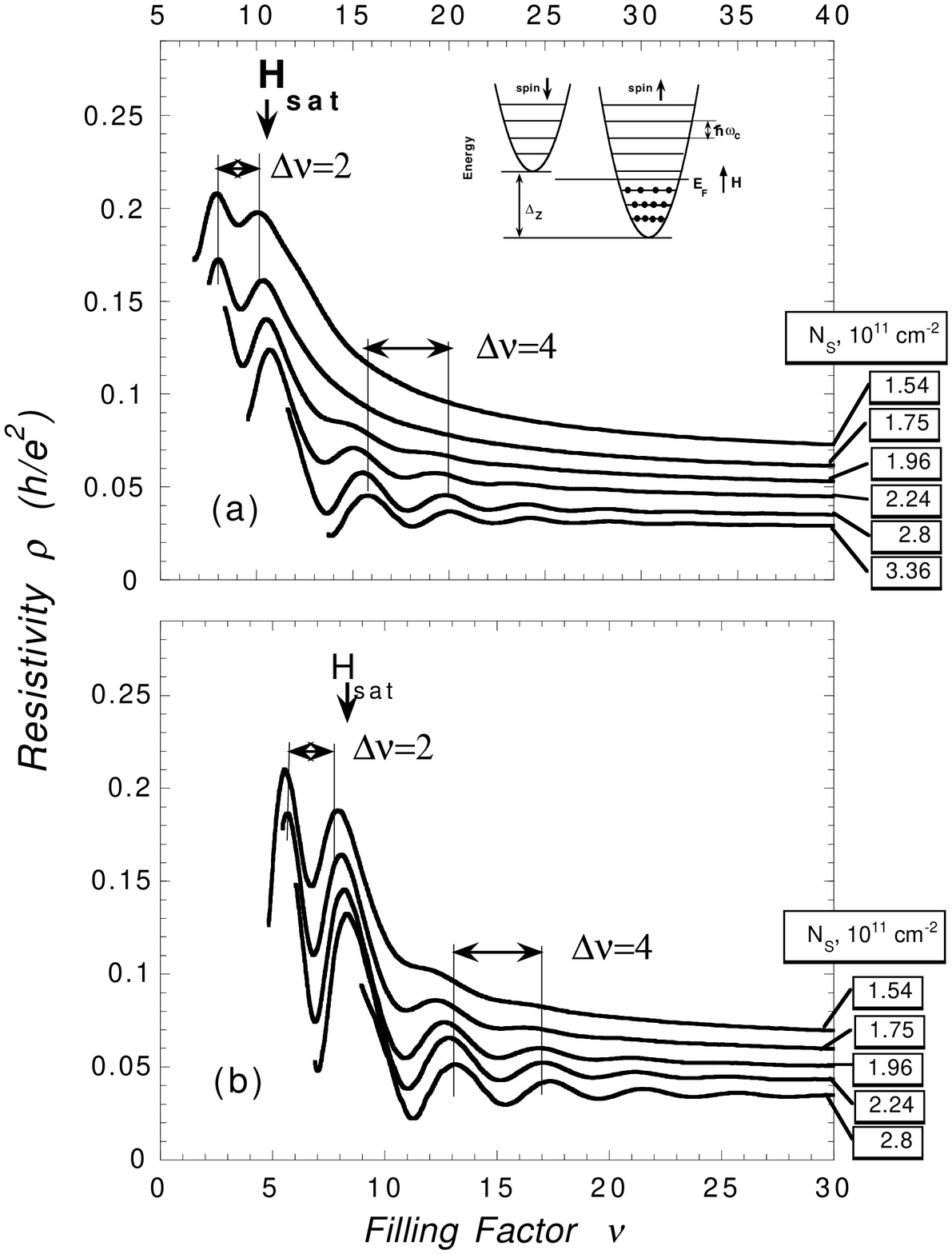} 
}
}
\refstepcounter{figure}
\parbox[b]{3.2in}{\baselineskip=12pt \egtrm FIG.~\thefigure.
Resistivity $R_{xx}$ versus electron filling factor
$\nu=n_s \Phi_0/H_{\perp}$ showing SdH oscillations for
different
electron densities $n_s$, as labeled.  Data are shown
at $T=0.25 K$ for two different angles $\phi$ between the magnetic field
and the electron plane: (a) $\phi=6^o$; (b) $\phi=7.8^o$.
\vspace{0.10in}
}
\label{1}

For various different densities $n_s$, Figs. 2 (a) and 2 (b)
show
the resistance $R_{xx}$ vs filling factor $\nu$ at two different angles $\phi$
between the magnetic field and the 2D plane.  Similar curves were obtained at
other small angles.  The filling factor $\nu=n_s\Phi_0/H_\bot$ was
calculated using the relation $n_s=H_{\bot}/(R_{xy}ec)$.  The Hall
resistivity $R_{xy}$ and Hall coefficient $R_H$ were determined from the
low-field data, $R_{xy}=R_H\times H_{\bot}$,
({\it i. e.} in fields below the onset of quantum oscillations).

For the lowest densities
shown in Fig. 2 (a) at angle $\phi=6^o$, the system is in the high-field
saturated regime above $H_{sat}$ for filling factor $\nu<10$.  The large
arrow indicates
$H_{sat}$ for $n_s=1.54 \times 10^{11}$ cm$^{-2}$.
Quantum oscillations are clearly evident superimposed on the large plateau
value of $R_{xx}$ at small $\nu$.  In this region the period of the SdH
oscillations corresponds to a change
in filling factor $\Delta \nu_{sat}=2$ (including the two-fold
valley degeneracy for a 2D layer of (100) silicon).  For higher filling factors
$\nu>10$ where the 2D electron system is below saturation ($H<H_{sat}$), the
period of the SdH oscillations is twice as long, namely $\Delta \nu=4$.
Similar behavior is shown for a bigger angle in Fig. 2 (b), where
the larger perpendicular component gives rise to stronger SdH  oscillations;
here the period doubling is found above $\nu \approx 8$.

The period $\Delta \nu=2$ of the oscillations at $H>H_{sat}$
corresponds to complete spin polarization of the electrons.
There are several scenarios that could account for a transition to full
polarization in
strong magnetic fields, depending on the nature
of the ground state of the system at $H=0$
\cite{NAS}, 
an issue that is currently under
debate.  Here we restrict the discussion to a simple model within a single-
particle description.  Using this approach we were able to explain the
doubling
of the frequency of the SdH oscillations at $H>H_{sat}$.  However,
the detailed behavior of the oscillations in small perpendicular fields is not
fully consistent with this model.

As shown schematically in the inset to Fig. 2, the spin-up
and spin-down electron bands are split by the Zeeman energy
$\Delta_Z=g\mu_B H$, while
the spacing between the Landau levels,
$\hbar \omega_c= \hbar eH_{\bot}/mc$, is determined by the perpendicular
component of the field.  We consider the
progression as electrons are added to the system: for small densities,
$E_F<\Delta_Z$, electrons are added to Landau levels in
the spin-up band only, corresponding to a SdH periodicity $\Delta \nu=2$
(including a factor 2 for the valley degeneracy in silicon); at
high
densities, $E_F>\Delta_Z$, twice as many electrons are required
to fill both spin-up and spin-down Landau levels, yielding the double
period, or $\Delta \nu=4$\cite{Elihu}.  An equivalent
argument holds for fixed density as one reduces the magnetic field.  Thus, the
shorter period $\Delta \nu=2$  at $H>H_{sat}$ signals the onset of full
polarization of
the electron system\cite{splitting}.  The relationship between $H_{sat}$ and
complete spin polarization 
was also found by Okamoto {\it et al}
\cite{okamoto} using a different experimental method.

The observed period-doubling is consistent with this simple model
only if the spin-up and spin-down levels are degenerate or nearly so, so that
$\alpha=\Delta_Z/ \hbar \omega_c = i$ with $i$ an integer.
The double period should revert to a single period when
$\alpha=i+1/2$, corresponding to a spin-up Landau level between two
spin-down Landau levels.  The ratio
$\alpha$ can be varied experimentally by changing the angle $\phi$,
or by using the fact that the electron-electron interaction-enhanced g-factor
(and thus $\Delta_Z)$ decreases with increasing electron density in silicon
MOSFETs\cite{okamoto,Ando}.  By taking data over a broad range of densities,
we were able to smoothly vary $\alpha$ by more than $1$.  Close examination
of the data shows that the double period in fields below $H_{sat}$,
although stable over a broad range, does break down in a narrow
range of densities that is different for different angles $\phi$, as expected
within this model.

SdH oscillations reflect changes in electron scattering due to periodic
oscillations of the density of states at the Fermi level\cite{abric}.  
For the weak perpendicular fields used in our expriments, there is strong 
scattering and the SdH oscillations are small\cite{Dingle}.  Unlike the 
situation that prevails in high magnetic fields, where the Landau levels are 
sharp and well-defined, the density of states is best 
represented in this regime by a harmonic expansion\cite{Ando}:
$$
D_{\downarrow, \uparrow}(E)=D_0(1+\epsilon \times cos(2\pi
(E \pm \Delta_Z/2)/\hbar \omega_c)  +O(\epsilon^2))  \eqno{(1)}
$$
Here $E$, $\Delta_Z=g\mu_B H$, and $\hbar \omega_c$ are the energy, Zeeman
energy and cyclotron energy, respectively.  The small parameter
$\epsilon=2 exp(-\pi/(\omega_c \tau)) \ll 1$\cite{Ando}
is proportional to the Dingle factor \cite{Dingle}.
Small variations in the resistivity are proportional to small variations
in scattering: $\Delta \rho/ \rho=\Delta W/W$.  Using the Born approximation,
$W \sim \int \delta (E-E_F)D(E)dE$, one can show that:
$$
\Delta \rho/ \rho= \cases {\epsilon \times cos(\pi \Delta_Z/ \hbar \omega_c)
cos(2\pi E_F^0/\hbar \omega_c)
, & $\Delta_Z <2 E_F^0
$\cr
\epsilon \times cos(4\pi E_F^0/\hbar \omega_c), & $\Delta_Z >2 E_F^0$}
\eqno{(2)}
$$
Here the Fermi energy $E_F$ is measured
from the bottom of the band at
$H=0$, $E_F^0$ is the Fermi energy at $H=0$, and $D(E)$ is the total density of
states: $D(E)=D_\uparrow(E)+ D_\downarrow(E)$.  We assumed $T=0$ and neglected
higher harmonic terms of order $\epsilon^2$ in Eq.(1)
as well as higher order corrections due to oscillations
of the Fermi energy.  This demonstrates that the SdH period
changes by a factor of 2 when $\Delta_Z >2 E_F^0$, corresponding to
full polarization of the electrons.

The term
$A = \epsilon cos(\pi \Delta_Z/ \hbar \omega_c)$ of Eq. (2) depends
on the ratio $\alpha=\Delta_Z/\hbar\omega_c$, which is fixed for a given angle
and electron density and does not
vary with magnetic field.  It determines the overall amplitude of
the oscillations at fields below $H_{sat}$, when $\Delta_Z <2 E_F^0$.  This
amplitude has a maximum when $\alpha=\Delta_Z/ \hbar \omega_c=i$ is an
integer, corresponding to spin
up and spin down densities of states oscillating in phase, and vanishes
when $\alpha = i + 1/2$ (see Eq.1).  Figure 3 shows data over a
narrow region near $\alpha=i+1/2$ where the amplitude of the first
harmonic is small, allowing detailed examination of higher harmonic terms
(see Eq.(1)).

Based on the usual assumption that the SdH oscillations are determined by the
total density of states, $D(E)=D_\uparrow(E) + D_\downarrow(E)$, one expects
and generally observes\cite{example}
the progression illustrated schematically in Fig. 4: a minimum which becomes
progressively deeper develops at the center of each maximum (see curves (a)
and (b)), gradually splitting it into two separate maxima (curve (c)).
Thus, the minima of the second harmonic (curve (c))
are at the positions of the maxima of the first harmonic (curve (a) in 
Fig. 4).

\vbox{
\vspace{0.2in}
\hbox{
\hspace{-0.2in} 
\epsfxsize 3.3in \epsfbox{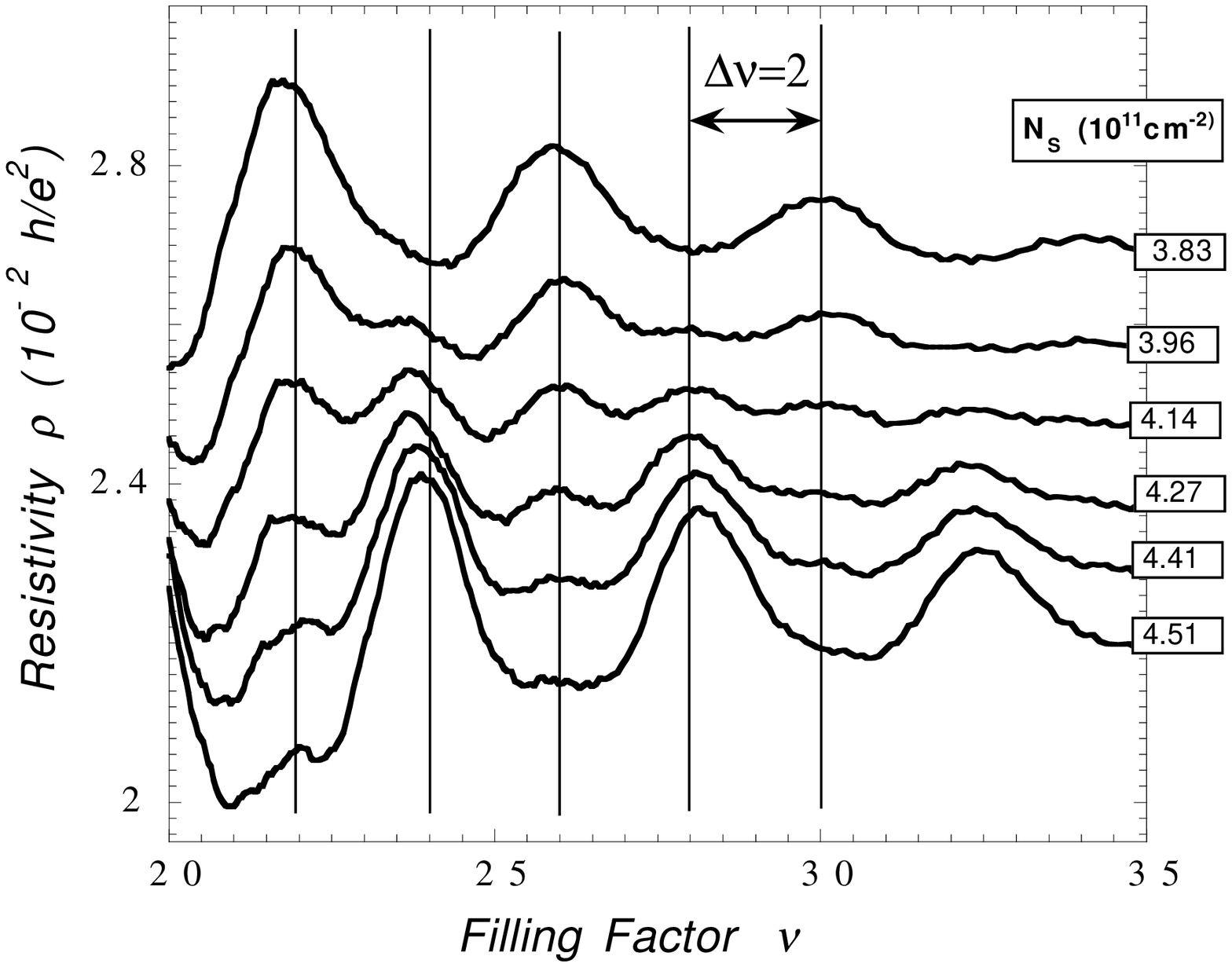} 
}
}
\parbox[b]{3.3in}{\baselineskip=12pt \egtrm FIG.~\thefigure.
In magnetic fields below saturation, $H<H_{sat}$, the SdH
oscillations are shown for a narrow range of electron densities near
$\alpha =\Delta_Z/ \hbar \omega_c = g\mu_B H/\hbar \omega_c = (i+1/2)$; due to
the $e-e$
enhancement of the g-factor, each density corresponds to a slightly
different value of $\alpha$.
Note that the maxima of the second harmonic
($n_s=4.14 \times 10^{11}$ cm$^{-2}$) are in phase with maxima of the first
harmonic($n_s=3.83$ and $4.51 \times 10^{11}$ cm$^{-2}$).
\vspace{0.10in}
}
\label{3}

\vbox{
\vspace{0.2in}
\hbox{
\hspace{-0.2in} 
\epsfxsize 3.3in \epsfbox{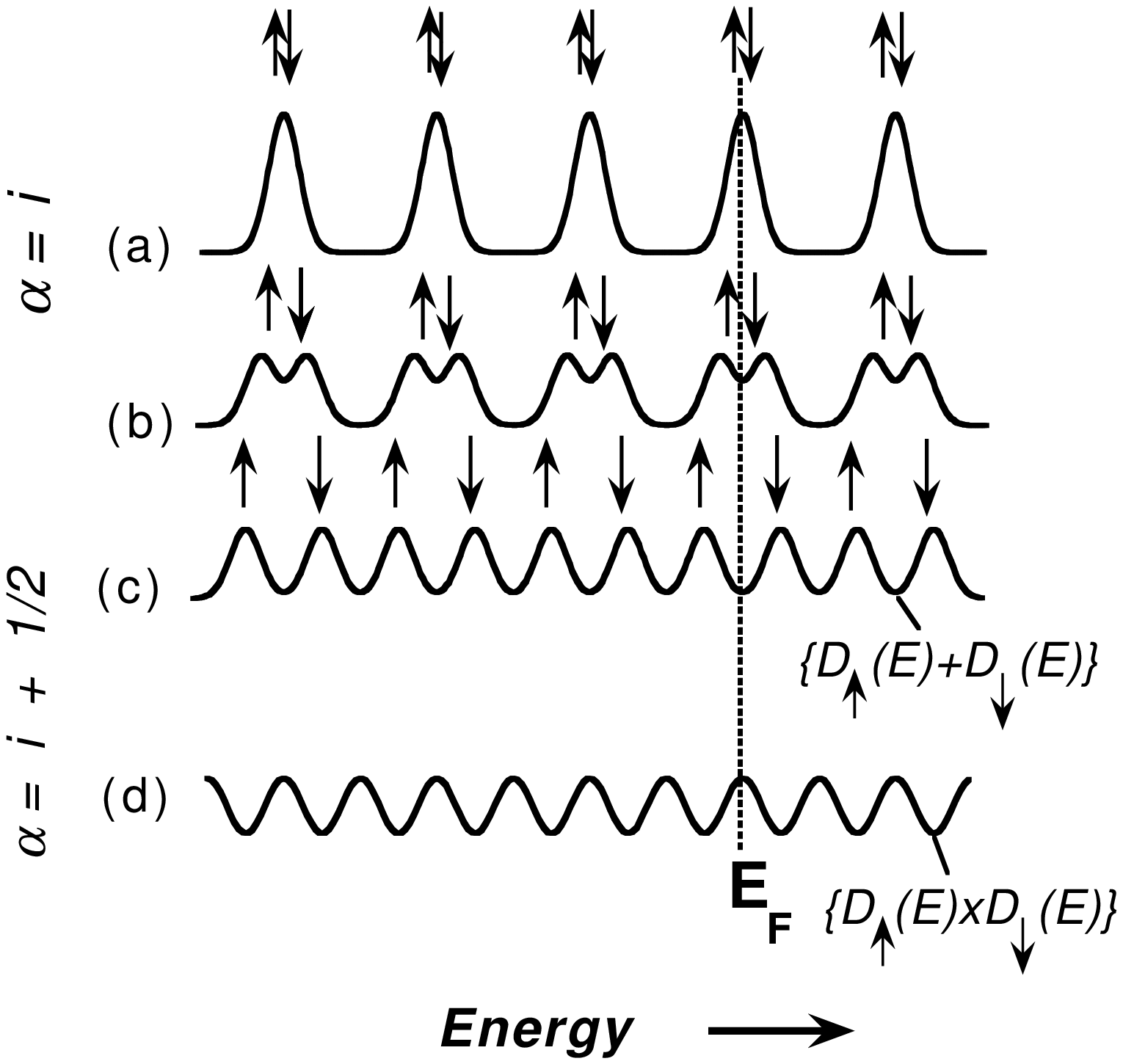} 
}
}
\parbox[b]{3.3in}{\baselineskip=12pt \egtrm FIG.~\thefigure.
Schematic of SdH oscillations for a density of states
$D(E) = D_\uparrow(E) + D_\downarrow(E)$ when: (a) the spin-up and spin-down
Landau
levels are degenerate, $\alpha =\Delta_Z/ \hbar \omega_c = i$; (b)
$\alpha = i+\delta$ for small $\delta$; and (c)
$\alpha = i+1/2$. 
Note that the maxima in
(a) and (c) are $180^o$ out of phase.  Curve (d) illustrates maxima for the 
second harmonic in phase with maxima of the first obtained from assuming 
a scattering probability of the form 
$W =f(D_\uparrow(E)\times D_\downarrow(E))$, in agreement with the experimental 
behavior shown in Fig. 3.
\vspace{0.10in}
}
\label{3}

However, careful examination of the data of Fig. 3 shows that the behavior 
below $H_{sat}$ observed experimentally in silicon MOSFETs is quite different: 
no minima develop within 
the maxima, splitting them into
two; instead, the maxima simply diminish in amplitude and new neighboring 
maxima appear and grow in amplitude.  The maxima of the first 
($n_s=3.83$ and $4.51\times 10^{11}$ cm$^{-2}$)
and second ($n_s=4.14\times 10^{11}$ cm$^{-2}$)
harmonics are in phase (as in curves (a)
and (d) of Fig. 4) rather than $180^o$ out of phase.  The origin of this
unusual behavior is not clear and warrants further careful study.
Interestingly, the phase relation between first and second harmonics observed 
in our experiments can be obtained within the single particle model used 
earlier if one considers the product of spin-up and spin-down densities of
states, $W =f(D_\uparrow(E)\times D_\downarrow(E))$, rather than their sum; 
curve (d) of Fig. 4 is the result of such a calculation.  
This suggests there is a sizable contribution to the electron scattering from
events that depend on the overlap of spin-up and spin-down densities of
states,
perhaps reflecting enhanced scattering of electrons of opposite spin.

In summary, measurements of small-angle Shubnikov-de Haas oscillations indicate
that the period of the oscillations changes by a factor of two at the
magnetic field $H_{sat}$ above which the resistance has reached saturation.
We attribute the abrupt change in period to
the onset of full polarization of the electron spins.  The period
doubling in fields below $H_{sat}$ is stable with
respect to the angle between the magnetic field and the 2D plane, and
is observed for all electron densities
except in a narrow interval, where the amplitude of the first harmonic of
the SdH oscillations vanishes and the second harmonic is observable.
The phase observed for the second harmonic relative to the first
is consistent with SdH oscillations due to scattering
events that depend on the overlap instead of the sum of the
spin-up
and spin-down densities of states at the Fermi level.  This unusual
behavior may reflect the importance of many-body interactions in the 2D system.

We are grateful to S. Bakker and R. Heemskerk for their contributions in
developing and preparing the MOSFETs used in this work.  We thank E. Abrahams,
L. Ioffe, F. Fang, A. Fowler, S. V. Kravchenko, A. Shashkin, X. Si,
S. Chakravarty, D. Schmeltzer, F. Stern, M. Raikh and U. Lyanda-Geller for
illuminating discussions.  We are grateful to A. Shashkin and
S. V. Kravchenko for valuable
comments on the manuscript.
This work was supported by U.~
S. Department of Energy grant No.~DE-FG02-84ER45153.

\end{multicols}


\begin{references}
\bibitem{alignment} Parallel alignment was determined by choosing the angle
for which $R_{xy}=0$.
\bibitem{defineH} $H_{sat}$ was chosen as the magnetic field where the
resistivity has reached 90\% of its constant high-field value.
\bibitem{splitting} An alternative explanation is that each Landau level
splits into spin-up and spin-down
sublevels in high magnetic field.  It is unlikely that this
would occur in every case at precisely the magnetic field where the resistance
reaches saturation.  Moreover, one expects the Landau levels to be
broader for the larger resistances at high fields, making such splitting
less likely.  Also at $H>H_{sat}$ we do not see any interference effects
corresponding to two spin bands.
\end{references}
\end{document}